\documentclass[aps,prl,twocolumn,color,psfig,epsf]{revtex4}
\usepackage{amssymb}
\usepackage{graphicx}
\usepackage{color}
\usepackage{amsmath}

\begin{document}

%\begin{frontmatter}



\title{Oscillation in the SIRS model} %% Article title


\author{Davide Marenduzzo} %% Author name
\author{Aidan T. Brown} %% Author name
\author{Craig W. Miller} %% Author name
\author{Graeme J. Ackland}

\affiliation{School of Physics and Astronomy, University of Edinburgh, Edinburgh, EH9 3FD, UK}

%% Abstract
\begin{abstract}
We study the SIRS epidemic model, both analytically and on a square lattice.  
The analytic model has two  stable solutions, post outbreak/epidemic (no infected, $I=0$) and  the endemic state (constant number of infected: $I>0$).  When the model is implemented with noise, or on a lattice, a third state is possible, featuring regular oscillations.  This is understood as a cycle of boom and bust, where an epidemic sweeps through, and dies out leaving a small number of isolated infecteds. As immunity wanes, herd immunity is lost throughout the population and the epidemic repeats.  The key result is that the oscillation is an intrinsic feature of the system itself, not driven by external factors such as seasonality or behavioural changes.  The model shows that non-seasonal oscillations, such as those observed for the omicron COVID variant, need no additional explanation such as the appearance of more infectious variants at regular intervals or coupling to behaviour.  We infer that the loss of immunity to the SARS-CoV-2 virus occurs on a timescale of about ten weeks.
\end{abstract}

\maketitle

\section{Introduction}
\label{intro}

%% Use \subsubsection, \paragraph, \subparagraph commands to 
%% start 3rd, 4th and 5th level sections.
%% Refer following link for more details.
%% https://en.wikibooks.org/wiki/LaTeX/Document_Structure#Sectioning_commands

Since the COVID epidemic there has been an increased interest in computer simulation of epidemics, using both detailed and complex models.  The most widely-used approach is the so-called compartmental models in which the population is divided into ``compartments'' according to their disease state.  The best known is the Susceptible-Infected-Resistant (SIR) model from which follow some fundamental ideas about epidemics.

\begin{itemize} 
\item Cases rise exponentially.
\item There is a balance between infectiousness and recovery rate below which an infection will naturally die out;   
\item After enough people have been exposed or vaccinated, herd-immunity is attained so that cases decline exponentially or faster.
\end{itemize}

The SIR model was introduced in 1927 in the wake of the Spanish flu pandemic by Kermack and McKendrick\cite{kermack1927contribution}. Its inadequacy in describing the endemic state was recognised quickly, and addressed by the same authors by introducing a steady flux of susceptible individuals into the system\cite{kermack1932contributions} or considering partial immunity\cite{kermack1933contributions}. Perhaps the key discovery in that work was the existence of a threshold beyond which the epidemic can spread throughout the population at an exponential rate.

The exponential growth rate, which is driven by the so-called reproduction number $\mathcal{R}$, defined here as the mean number of new infections that each infected causes, became totemic as an indicator of the severity of the COVID epidemic, with weekly public reports issued by government committees\cite{manley2024combiningR,manley2024combining,vegvari2022commentary} and high values triggering public interventions. 
Early models of the epidemic, particularly the Imperial College ``Report 9'' \cite{ferguson2006strategies,ferguson2020report,report9input,flaxman2020estimating} based on the highly detailed CovidSim model\cite{githubcovidsim}, were instrumental in backing the UK lockdown policy.  However, despite its complexity in modelling interactions on a network describing the entire UK, the results of this model were essentially similar to a simple SIR compartmental model whose key initial parameter was based on fitting $\mathcal{R}(t)$-numbers\cite{githubcovidsim,codecheck,rice2020effect} back-derived from the observed rate of increase of cases, assumed exponential. Overall, despite the high complexity and detail of the models used\cite{kerr2021covasim}, the main results can be understood in relatively simple mathematics \cite{kucharski2020effectiveness,thompson2020key,gog2021vaccine,thompson2021sars}.

In the initial stages of the epidemic there were ``waves'' of infections, each suppressed by external actions such as lockdowns, and recurring once the lockdowns were lifted.  After the omicron variant arrived, the combination of its high infectivity and reduced lethality led to the gradual phasing out of legal restrictions and an acceptance in the UK that SARS-CoV-2 is endemic. The omicron variant of SARS-CoV-2 is now in an endemic state worldwide, yet the SIR model does not describe an endemic state.  The simplest extension to SIR is to add the possibility of recovered individuals becoming susceptible again - the so-called SIRS model, also known as ``forest fire''\cite{bak1990forest,drossel1992self}.  This model does support an endemic state. The mechanism for this ``waning immunity'', which allows recovered to susceptible transitions is not specified, and may come either from the host immune system or from steady evolution of the virus.

SIRS seems a better candidate for SARS-CoV-2 than an influx of new susceptibles 
%\textbf{but is anyone now proposing this? Aren't most people proposing new variants instead? AB }  Yes, but I want to address   Kermack and McKendrick even if nobody else has read it.
as proposed by Kermack and McKendrick\cite{kermack1932contributions}, first because the influx of new susceptibles is slow relative to the speed of the epidemic, and second because most people have now had more than one bout of COVID, indicating loss of immunity\cite{UKHSA-dashboard,ONS-dashboard}.

A surprising feature of the endemic SARS-CoV-2 omicron variant in the UK is that the number of cases went through eight distinct oscillations, with a regular period of about 10 weeks, up to when UKHSA records ceased publication in 2023 \cite{UKHSA-dashboard,ONS-dashboard} and similar cessations worldwide\cite{worldometers}.  Reporting of absolute case numbers is unreliable, but it is clear that the peaks exhibited some 300-400\% more cases than the troughs. Anecdotally, this oscillation seems to have continued since monitoring stopped.%~\textbf{Does this include the 10-week period? AB} Yes.

This simple oscillatory behaviour happened in the absence of external interventions and with minimal public awareness. It strongly suggests that the oscillations are intrinsic to the disease, and any models of COVID should reproduce it. However, to our knowledge, none of the panoply of detailed models deployed worldwide during the pandemic predicted a steadily-oscillating endemic state. It is therefore of interest to understand what type of model is required to produce such an oscillating endemic state. 

It is, of course, possible to fit behaviour with additional parameters in a model. 
The simplest method is to introduce an external periodic driving force: the system will then oscillate at that frequency.  In the case of epidemics, one possibility is to make the infectivity of the disease - a parameter $\beta$ - seasonally varying.  Alternately, changes in $\beta$ can be triggered by reaching infection thresholds, imitating non-pharmaceutical interventions such as lockdowns, which are introduced to suppress an epidemic~\cite{ferguson2020report}. 
There are numerous studies introducing additional terms and factors to the SIRS model\cite{cooke1996analysis,cai2015stochastic,cai2017stochastic,satsuma2004extending,greenhalgh2003sirs,lahrouz2011global,federico2024optimal,kaur2014modeling,kabir2019analysis,zhao2014threshold,chang2017analysis,davidson2008effects,el2024sirs}
or running the model on mathematically-defined or empirical complex networks\cite{li2014analysis,pare2020modeling,colizza2006modeling,colizza2008epidemic,valdano2015analytical,colizza2006role,zhang2018dynamics,van2004emergent}. It is also interesting to ask whether oscillations can arise in relatively simple spatial or non-spatial versions of the SIRS model, without requiring any external forcing, behavioural feedback, additional concepts or parameters. Sustained oscillations in the SIRS model have been demonstrated in the pair approximation, which is non-spatial, but involves tracking the number of S-S, S-I... contacts~\cite{rozhnova2009}, and on small-world networks~\cite{kuperman2001}.

\begin{figure*}
\includegraphics[width=\textwidth]{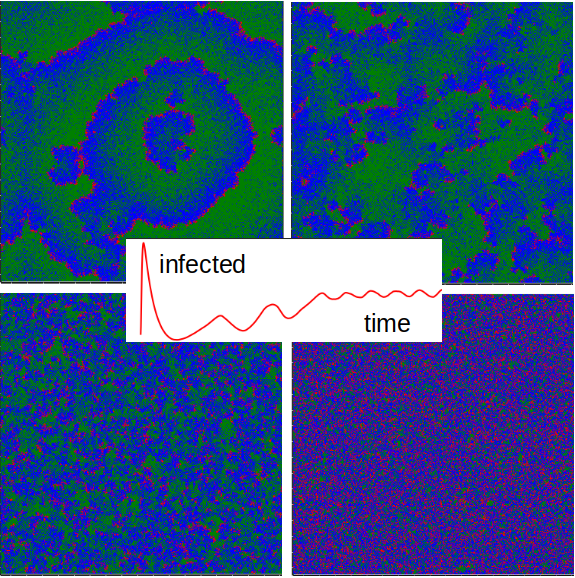}
\caption{Images of SIRS on a lattice, with SIR coloured Green,Red,Blue.  the first three show a clear separation of timescales ($\beta,\gamma,\delta)=(0.88,0.1,0.007)$. At short times the epidemic has a well defined, single wave structure spreading through the system, with secondary waves initiating and merging once immunity is lost.  At longer times the endemic state has outbreaks of a characteristic size.  With more rapid immunity loss, the oscillation frequency and spatial correlation is reduced ($\delta=0.014$, bottom left) and eventually lost ($\delta=0.04$, bottom right). \label{graphicalabstract}}
\end{figure*}

In this paper we study the relationship between oscillations that arise in simple non-spatial and spatial SIRS models (Fig.~\ref{graphicalabstract}). We compare noise-driven oscillations in the mean-field model, and oscillations on a square lattice, and show that these different realizations of the SIRS model give rise to qualitatively different scalings with the controlling parameters of the model.

%Rather than postulate another mechanism, we here demonstrate that oscillations can occur spontaneously in SIRS models without requiring any external forcing, behavioural feedback, additional concepts or parameters.

\section{Methods}

\subsection{Deterministic ODE SIRS model}
Compartmental models are the staple methodology of epidemic modelling.  In the case of SIRS, there are three compartments, Susceptible (S) Infected (I) and Resistant (R).    Transitions are possible from S to I,  from I to R, and from R to S.  The model is parameterised by three rates, one for each transition. In the mean-field approach, the SIRS model can be written as a set of nonlinear differential equations
\begin{eqnarray}
\frac{dS}{dt} & = & -\beta IS + \delta R \,, \nonumber \\
\frac{dI}{dt} & = & \beta IS - \gamma I\,,\nonumber\\
\frac{dR}{dt} & = & \gamma I - \delta R \,.\label{eq:SIR}
\end{eqnarray}

We note that the total population is conserved, so that we can write the populations as fractions
\[ S+I+R=1 \,,\]
and reduce the model to two equations in two unknowns
\begin{eqnarray}
\frac{dS}{dt} & = & -\beta IS + \delta (1-S-I)\,, \nonumber\\
\frac{dI}{dt} & = & \beta IS - \gamma I\,,
\label{eq:frac}
\end{eqnarray}
with a typical initial state for studying an outbreak of $I\gtrsim 0$, $R=0$. Here, $\beta,\gamma,\delta$ are rates of transfer between compartments.

SIRS is essentially a {\it {flow}} model where individuals cycle through the compartments.  It does not satisfy detailed balance so concepts borrowed from equilibrium statistical mechanics should be applied with caution.  In particular, there is no flow $I\rightarrow S$ which means there may be non-equilibrium solutions with a non-zero current, or cyclic flow.
These equations allow for two steady state solutions, the trivial $I=0$ and $I>0$, which is known as the endemic state.

By setting the time derivatives to zero, some simple algebra gives results for the mean field steady state
\begin{align}
 S_\infty &= \gamma/\beta \,,  \nonumber\\
     R_\infty &= \gamma\frac{\beta-\gamma}{\beta\delta+\beta\gamma}\,, \nonumber \\
     I_\infty &= \delta \frac{\beta-\gamma}{\beta\delta+\beta\gamma} = \frac{\mathcal{R}_0-1}{\mathcal{R}_0 [1+\gamma/\delta]}\,, \label{eq:endemic}
\end{align}
where the first form for $ I_\infty$ emphasizes
that for small $\delta$, $ I_\infty$ scales with $\delta$, going to zero for $\delta=0$, which is the SIR model.  The second form introduces $\mathcal{R}_0$, the value of $\mathcal{R}$ 
in the limit $S\rightarrow $1, which is typically associated with $t=0$.

Although it appears to have three parameters, Eq.~\ref{eq:endemic} can be simplified by taking $\delta$ as the characteristic time.
This means that the behavior of SIRS is determined by just two parameters, for example, defining dimensionless variables $b=\beta/\delta$ and $g=\gamma/\delta$, leading to

\begin{align}
 S_\infty &= g/b \,, \nonumber\\
     R_\infty &=g\frac{1-g/b}{1+g}\,, \nonumber \\
     I_\infty &= \frac{1-g/b}{1+g}\,. \label{eq:endemic2}
\end{align}

Since one can only have fractions between 1 and 0, Eq.~\ref{eq:endemic} only has meaningful solutions for:
\[ \mathcal{R}_0 = \beta/\gamma=b/g>1\,.\]
 The sign of $\mathcal{R}_0-1$ determines whether the initial behaviour is exponential growth or decay, and whether an epidemic can occur from a single initial case. $\mathcal{R}_0$ is understandable as the number of people each infected person can infect in the limit of $S\rightarrow 1$. By definition, endemic behaviour requires $I_\infty > 0 $, which is satisfied for all $\mathcal{R}_0> 1$ (Eq.~\ref{eq:endemic}).  One cannot have negative numbers of people, so Eq.~\ref{eq:frac} also requires $R_\infty< 1$.
%A mathematical solution for $R_\infty=1$ exists, but involves a negative $\beta$ so is of no physical consequence.

The mean-field SIRS model does not have an oscillatory endemic solution, although it is possible for the system to undergo damped oscillation en-route to the endemic state. It is, of course, possible to achieve sustained oscillations, by introducing additional effects. It is therefore interesting to ask whether and how long-term oscillations can be introduced without additional parameters.

\subsection{Stochastic ODE SIRS model}

In a number of ecological and epidemiological systems whose deterministic behaviour shows damped oscillations, the addition of white noise produces sustained oscillations by preventing the system from ever reaching its fixed point: this is so-called stochastic resonance\cite{mckane2005predator,alonso2007stochastic,black2010stochastic,black2012stochastic,perez2023universal,benzi1982stochastic}. To examine this in the SIRS model we run numerical calculations of Eq.~\ref{eq:SIR} with each parameter $\beta,\gamma,\delta$ multiplied by $\eta(t)$, a white-noise random variable between $1-\eta_{max}$ and $1+\eta_{max}$, at every timestep $dt$ of  the integration.  In code, this means transferring some fraction of the population
between compartments according to: 
\begin{eqnarray}
S\rightarrow I   = \beta IS\eta_\beta(t) dt \\
I\rightarrow R   = \gamma I\eta_\gamma(t) dt \\
R\rightarrow S   = \delta R\eta_\delta(t) dt \,.\label{eq:SIRstoch}
\end{eqnarray}
here, we set $\eta_{max}=1$ in all cases unless stated otherwise.

Both these ODE models are mean-field models, insofar as there is no memory of previous states or correlation between I and S individuals or - by implication, an infected person will never encounter their infector again. The choice of multiplicative noise means the populations can never go to zero\cite{parsons2024probability}

\subsection{SIRS on a lattice}
An alternative realisation of SIRS comes from introducing a spatial dimension or network wherein each individual is connected to only a few others. 
The easiest way to implement this is as a stochastic cellular automaton lattice model. In such a model, one creates a network of sites, each of which represents an individual and has a state $S,I,$ or $R$.  The infection term $\beta IS$ then operates only between two connected sites.  Most detailed epidemic models are of this type, modified so that the network comes from a real geography and the infection probability ($\beta$) between any two sites depends on the location, age, behaviour etc. of those individuals, either postulated or obtained from survey data\cite{edeling2021impact,leung2021quantifying,githubcovidsim,ferguson2020report,rice2020effect,kerr2021covasim}.  It is nowadays possible to run such models with many millions of sites, mimicking the population of a country: here we have run our python SIRS code, with animation, on an $L\times L$ square grid with 10$^8$ sites using a single CPU.

One effect of introducing a spatial dimension to this type of model is to generate waves which spread through the system~\cite{van2004emergent}, similar to the well-known continuum FKPP equation or KPZ equations\cite{fisher1937wave,kardar1986dynamic} which we have found in a number of applications\cite{ackland2007cultural,cohen2012boundaries,ackland2014cultural}. 
In a sufficiently large system the wave will spread at a fixed rate, such that the number of infecteds grows at best as a power law - giving an apparent $\mathcal{R}$-number of 1, if one attempts to derive $\mathcal{R}$ from fitting an exponential to the growth rate \cite{ackland2022fitting}.  In the limiting case of $\delta=0$ the SIRS model becomes equivalent to SIR, for which on a lattice the epidemic threshold  is
$\beta/\gamma>2$. The factor of 2 reflects that for the wave to advance each I site must infect, on average, at least one other site {\it immediately ahead of the wave}. This can be contrasted with the mean field case for which  each I site must infect, on average, at least one other site {\it anywhere in the system} (as no spatial effects are included at that level).

A typical approach in modelling is to use periodic boundary conditions.  For a sufficiently large system, a wave solution will traverse the system at some speed $v$, giving rise to oscillations at frequency $L/v$.  The system-size dependence of the period oscillation is a clear sign that this type of oscillation is an artifact of the boundary conditions\cite{heiba2018boundary}.

However, the spatial dimension can cause more authentic oscillations where a resource (here S) is overconsumed by some population (here I),  leading to an environmental collapse and subsequent population collapse.
If the resource regenerates on a slow timescale (here $\delta$), while the population survives only at a very low level 
then a second epidemic can emerge.  Such simulations use stochasticity,  but unlike stochastic resonance it is not the noise which amplifies the oscillations.   This type of ``boom-and-bust'' behaviour has been reported in economic and ecological models\cite{ackland2007strategy,ackland2014cultural,matsuda1992statistical,provata1999oscillatory}

Here we demonstrate that the SIRS model on a lattice exhibits this boom-and-bust style oscillation. The model is implemented as a cellular automaton wherein the autonomes have three states. The model has only three parameters, one of which simply sets the timescale and does not affect the qualitative behaviour.
Transitions between the states are implemented by a stochastic cellular automaton which implements the same concept as the differential equations.  The algorithm is as follows:

\begin{itemize}
    \item A site is chosen at random
    \item if it is in state Infected, it becomes Resistant with probability $\gamma_a$  
    \item if it is in state Resistant, it becomes Susceptible with probability $\delta_a$
    \item if it is in state Susceptible, it becomes Infected with probability $\beta_a/4$ if any of its neighbours is Infected.
\end{itemize}

Here the $a$ subscript denotes ``algorithm''.
We note that this is not a unique way to represent the SIRS model - in particular infections occur after the choice of the infected site S.  One could write an algorithm where infection occurred based on a random choice of an I-S link.  Other networks are possible, and one can write a version equivalent to the mean field case with all sites connected to all others.

\subsection{Relationship between the parameters}

Compared to the ODE form, $\gamma_a$ and $\delta_a$ are exactly equivalent exponential decay rates, sampled stochastically rather than deterministically.  To compare with the $\beta IS $ term, the algorithm implies that 
\begin{itemize}
\item S is selected at random, so the event rate is proportional to S

\item $\beta_a$ sets the infection probability as an exponential decay rate

\item ``If any neighbour is I'' -- in the mean-field limit, if $I$ is the fraction of Infected, this condition is met with probability $1-(1-I)^4$ which is $4I$ in the low-$I$ limit.
\end{itemize}

In comparing models, it is important to understand how the parameters are related.  We do this by comparing the limit of $S\rightarrow 1$, which is the typical ``patient-zero'' starting case for an epidemic. The exponential decays of recovery and loss of immunity are the continuum limit of the Poisson process in the algorithm, so $\gamma_a=\gamma$ and $\delta_a=\delta$. The factor of 4 in the algorithm compensates for there being four neighbours, such that $\beta_a=\beta$ in the limit of $S\rightarrow 1$. The timescales of the models are directly comparable: in the cellular automaton each ``timestep'' -- or ``sweep'' corresponding to a number of $N$ attempted updates according to the algorithm just described, where $N$ is the number of sites on the lattice --  can be considered of order one day. %~\textbf{I'm a little unsure what is meant by a timestep. According to the algorithm there is only one site selected each time step, so on an $N$-site lattice each person would have to wait $N$ days before interacting with their neighbours. So does a timestep mean $N$ site selections? AB Yes - in my code I update the whole lattice which leverages some speed-up via vectorization }. 

The numerical integration of the ODEs introduces an algorithm-related timestep, whilst the meaningful timescales are given by the inverse of the parameters $\beta,\gamma,\delta$.  For practical reasons, these times must be longer than the timestep for integration, whereas the lattice model requires probabilities $\beta/4,\gamma,\delta<1$.

\section{Results}

\subsection{Oscillations in SIRS}
Firstly, we calculate the frequency of damped oscillations in the mean field theory, by linearising the ODE system around the endemic equilibrium values (Eq.~\ref{eq:endemic2}), and taking the imaginary part of the complex eigenvalue. In terms of the dimensionless parameters $b$ and $g$ the equilibrium point is \( S_\infty = g/b\), \(I_\infty=(b-g)/[b(1+g)]\), and the Jacobian matrix around this equilibrium is:
%\[
\begin{equation}
A = \begin{pmatrix}
-\dfrac{b+1}{g+1} & -(g+1) \\
~\\
\dfrac{b-g}{g+1} & 0
\end{pmatrix}\,,
\end{equation}
%\]
whose eigenvalues are
\begin{equation}\lambda=\frac{-(b+1) \pm \sqrt{(b+1)^2-4(b-g)(g+1)^2}}{2(g+1)}\,. \label{eq:evals}\end{equation}

\begin{figure*}
    \includegraphics[width=\textwidth]{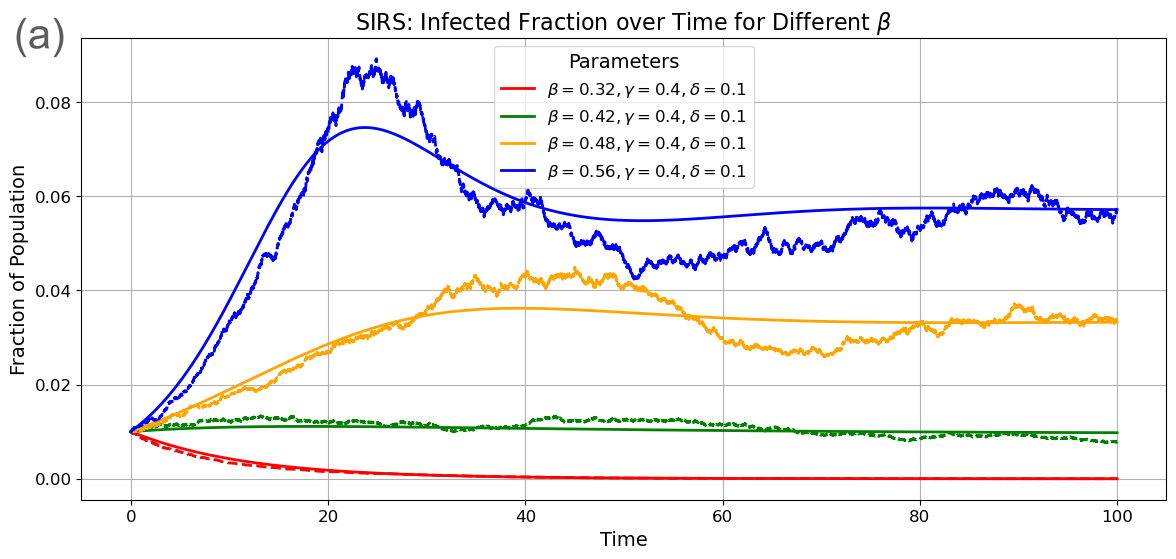}
    \includegraphics[width=\textwidth]{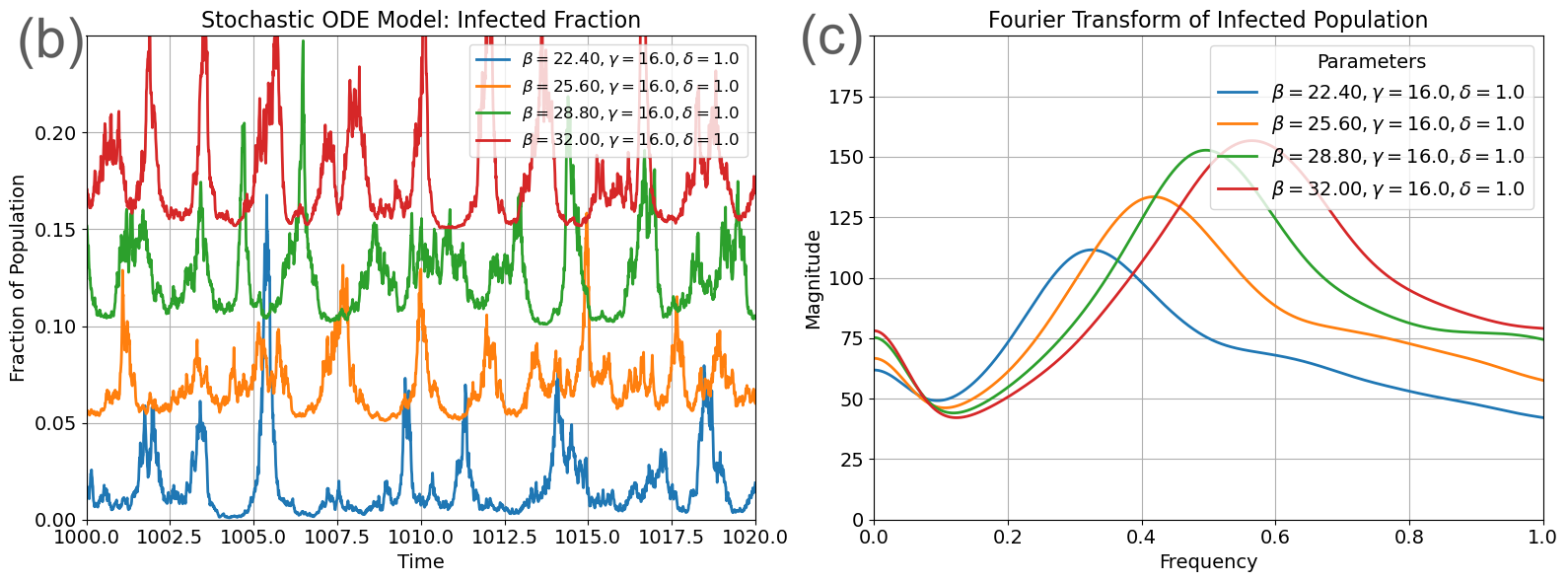}
    \caption{(a) SIRS model with various $\beta IS$, $\gamma I$, showing both deterministic and stochastic ODE models in initial stages of oscillating, endemic and non-endemic cases . (b) Section from endemic state of longer runs with stochastic dynamics. These ODEs are integrated with a timestep of 0.01.  Graphs of fractions of populations are displaced by 0.05 for clarity. (c) Fourier Transform of (b) showing the characteristic period. \label{fig:SIRS_dynamics}  
    }
\end{figure*}

It can be seen that ${\rm Re}(\lambda)<0$ for all $b$ and $g$, whereas there is a transition between a purely real $\lambda$, corresponding to exponential decay towards steady state, and a complex $\lambda$, corresponding to damped oscillations.
The imaginary part is illustrated in Figure~\ref{fig:phasediagram_MF}a.
In Figure~\ref{fig:SIRS_dynamics}  these two types of behaviour are demonstrated, along with the case $\mathcal{R}_0 > 1$ where the equilibrium is  $S=1$ and the infection is eradicated.

This transition between the oscillating and the non-oscillating state is shown in Figure~\ref{fig:phasediagram_MF}c and is at:
%\[
\begin{equation}
b=2(g+1)^2-1 \pm 2\sqrt{g(g+1)^3}\,.
\end{equation}
%\]

If we consider the epidemic-relevant scenario where waning immunity $\delta$ is the lowest rate, then $b,g\gg 1$ and the dominant term in the imaginary part scales as $\sqrt{b-g}$. Since both $b$ and $g$ are non-dimensionalised in relation to the characteristic timescale $1/\delta$,  the oscillation frequency scales as $\sqrt{\delta}$.

%is at (b+1)^2-4(b-g)(g+1)^2=0, which corresponds to b=2(g+1)^2-1 +/-2\sqrt{g}(g+1)^(3/2). This predicts a region near \mathcal{R}_0=1 where there are damped oscillations, and another region when \mathcal{R}_0\gg 1. I'm guessing that these would correspond to states where nearly everyone is infected (\mathcal{R}_0\gg 1) and almost no-one is infected (\mathcal{R}_0~1). In between, there is a large region without any oscillations. It looks like there is a bifurcation at b=1. I can show you some pictures and/or write this into the Overleaf if you want, but I should double-check the algebra first,

\begin{figure*}[t]%
\centering
\includegraphics[width=\textwidth]{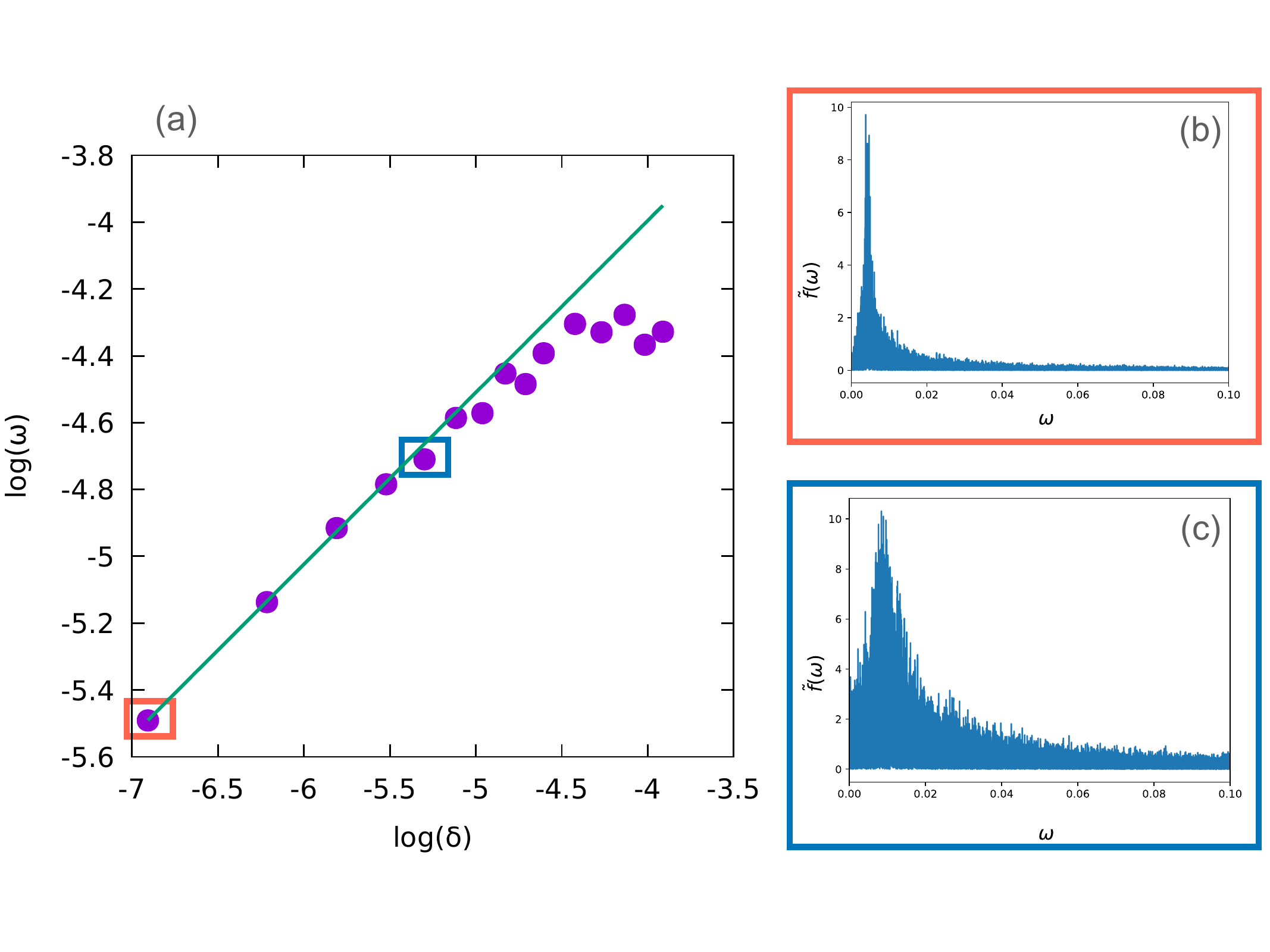}
\caption{Scaling of characteristic frequency in the stochastic SIRS ODE model  with waning of immunity $\delta$. (a) Typical frequencies were extracted from the peaks of Fourier transforms for selected time series showing oscillations for small $\delta$. The simulations were performed with fixed $\beta=0.8$, $\gamma=0.1$, and variable $\delta$. {\color{black}{The straight line shown was fitted in the range $-7\le \log{\delta}\le -5.5$ and}} corresponds to an exponent of $a\simeq 0.51$ (such that $\omega_{\rm max}\sim \delta^a$), compatible with square root behaviour for small $\delta$. (b-c) Examples of numerical Fourier Transforms (b: $\beta=0.8$, $\gamma=0.1$, $\delta=0.001$; c: $\beta=0.8$, $\gamma=0.1$, $\delta=0.005$). Throughout this figure, frequencies are given in units of inverse simulation time.}
\label{fig:I-freq}
\end{figure*}

The stochastic ODE model has similar behaviour initially to the deterministic one (see Fig.~\ref{fig:SIRS_dynamics}), however it exhibits stochastic resonance~\cite{benzi1982stochastic,mckane2005predator,alonso2007stochastic} with a characteristic frequency which can be extracted by Fourier transform of $I(t)$.  At very high $b$ and $g$ values we show that these are far from the harmonic oscillations implied by the eigenvalue analysis: rather, they comprise sharp peaks of infection separated by long inert periods.  The oscillations can be understood as follows: after an outbreak the number of $S$ states is highly reduced so that the effective R-number $\mathcal{R}(t)=S\beta/\gamma\ll 1$.  There then follows a period or order 
$\delta^{-1/2}$ during which immunity is lost and $S$ is replenished.  Eventually the epidemic threshold $\mathcal{R}(t)=S\beta/\gamma >1$ is reached, and another outbreak occurs, the size of which is highly sensitive to the noise in the system (and is zero without noise).
It can also be seen in Figure~\ref{fig:SIRS_dynamics} that there is some correlation between successive epidemics: larger ones are followed by a somewhat longer latent period.  There is no apparent correlation in the size of successive outbreaks.  In the low-$\delta$ limit, numerical integration shows that the frequency does indeed scale as $\sqrt{\delta}$ (Fig.~\ref{fig:I-freq}).  Away from this limit the behaviour is more complex and also depends on the size of the noise. 

\begin{figure*}[h]%
\centering
\includegraphics[width=0.6\textwidth]{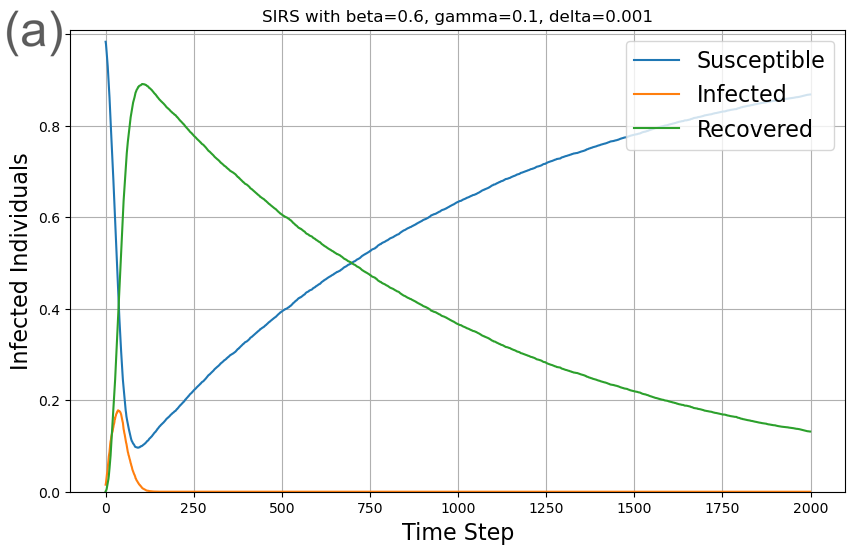}
\includegraphics[width=0.39\textwidth]{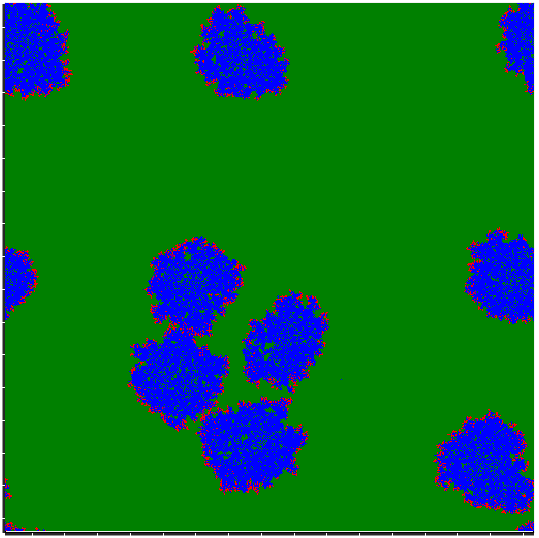}
\includegraphics[width=0.6\textwidth]{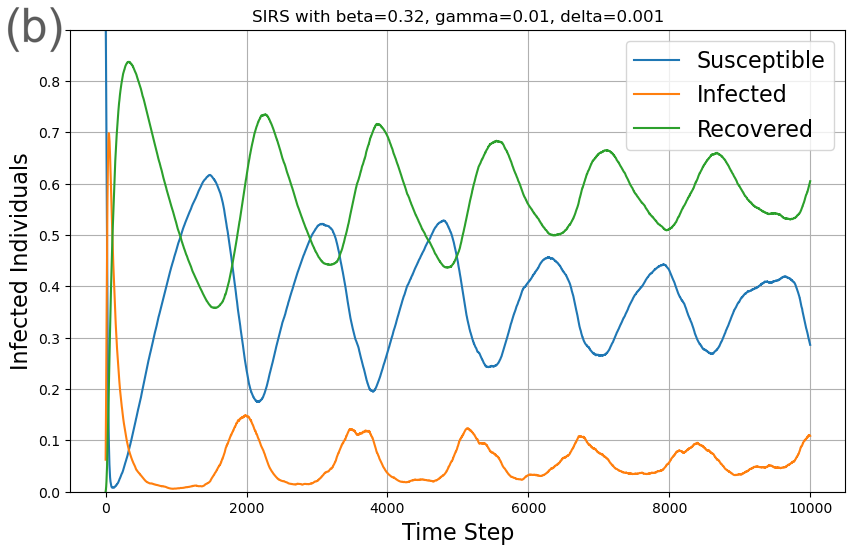}
\includegraphics[width=0.39\textwidth]{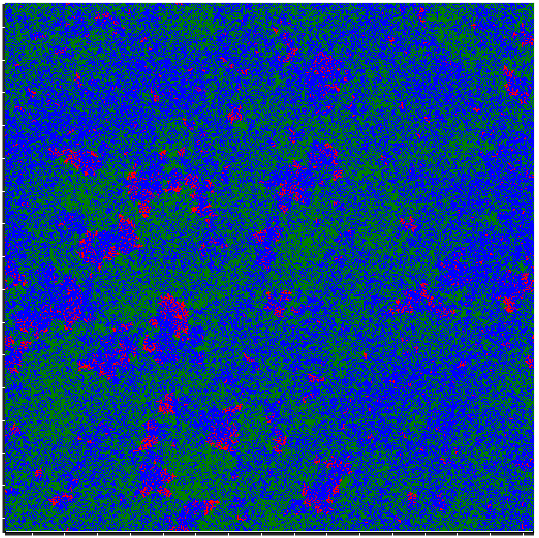}
\includegraphics[width=0.60\textwidth]{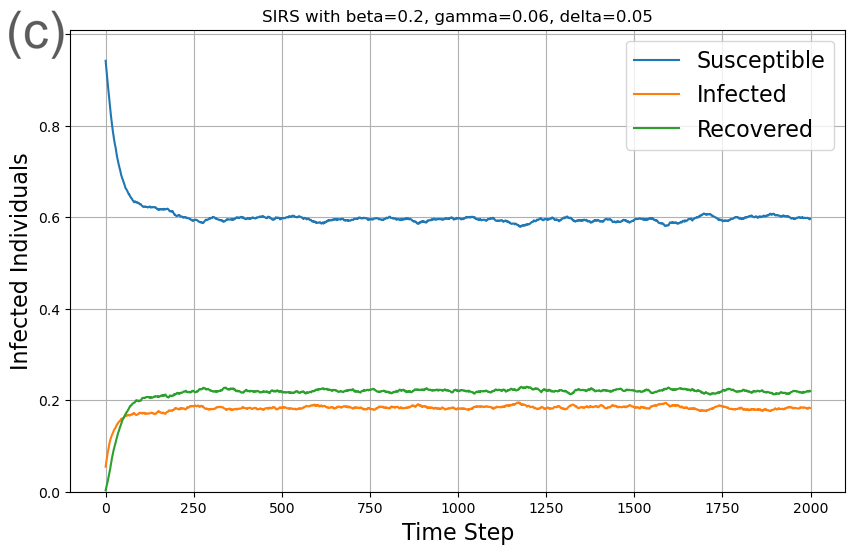}
\includegraphics[width=0.39\textwidth]{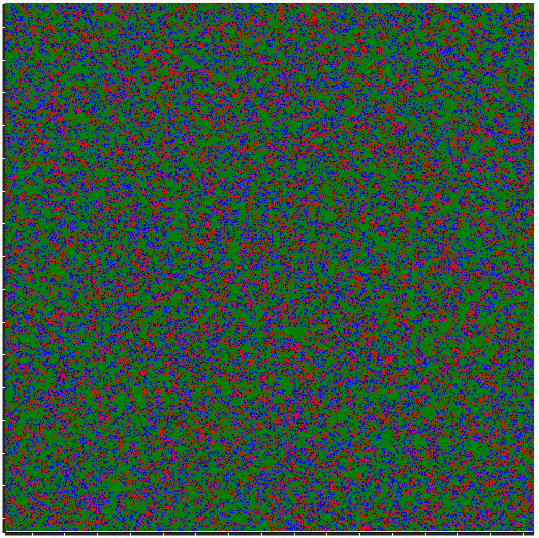}
\caption{Characteristic behaviours of the lattice model.
Plot of number of sites infected as a function of time for representative cases, showing (a) epidemic, (b) oscillatory  and (c) endemic regimes, with representative snapshots of the grid showing S (green) I (red) and R (blue) sites.
Time is measured in units of lattice sweeps where each site attempts to update once.
%Also shown are the mean-field results with the same $\beta,\gamma,\delta$.
\label{fig:Itot-time}}
\end{figure*}

The lattice model also shows regimes of oscillatory and endemic behaviour. In Figure~\ref{fig:Itot-time} we choose representative trajectories of $I(t)$ to illustrate the three cases:  

Figure~\ref{fig:Itot-time}a) shows epidemic behaviour with $1< \mathcal{R}_0< 2$. Cases rise initially as in the ODE case, but the spatial correlations between R and I mean that a wave cannot be sustained, and the infection dies out. {\color{black} These parameters produce an epidemic which infects 30\% of the population.  It is quite striking how many more individuals are recovered compared to infected.}

Figure~\ref{fig:Itot-time}b) shows the oscillatory behaviour.  Here we have almost two orders of magnitude difference between infection rate and loss of immunity.  The periodicity is comparable to 1/$\delta$. {\color{black} As expected, the peak in susceptibility precedes the peak in infections, with the peak in recovereds still later.}

Figure~\ref{fig:Itot-time}c) shows typical endemic behaviour, where the loss of immunity is very quick, and reinfections are commonplace.

\subsection{Wave nature of the oscillations on a lattice}
It is instructive to image the system in order to understand the source of the oscillations. Figure~\ref{fig:Itot-time}d) shows a snapshot of the weakly oscillating system on a 500x500 system.  There are clear correlations between regions of S and R, with wavefronts of I sweeping between them, continually moving into the S rich region.  The frequency can be extracted by Fourier transform of I(t) {\color{black}{(Fig.~\ref{fig:I-freq_lattice})}}, and we find that, in the low $\delta$ limit, this is roughly proportional to $\delta$, in contrast with the $\sqrt{\delta}$ found in the eigenvalue analysis, {\color{black}{as well as in the stochastic ODE model}}.  

Figure~\ref{fig:Itot-time} shows snapshots of the time variation of the spatial model starting from an initial ``patients-zero'' scenario.  The first wave of SIRS infection is an  SIR-type epidemic, reaching a much higher peak in $I$ than seen subsequently.  This is a single wave which essentially affects everyone in the system.  The second wave (top left in GA) grows into a partially-immune population, and is weaker and longer lived, but again (re)infects almost all sites. Subsequent waves lead to increasing amounts of $I(t)$ and are less correlated: essentially this means that the ``wavefront'' is becoming more convoluted and disjoint.  Again, each wave reinfects almost all sites.  For more rapid immunity loss (larger $\delta$), the frequency increases and the spatial correlation length is reduced (GA bottom left) until eventually the correlation length reaches the individual level and oscillations cease (GA bottom right).

\begin{figure*}[h]%
\centering%% For centre alignment of image.
\includegraphics[width=\textwidth]{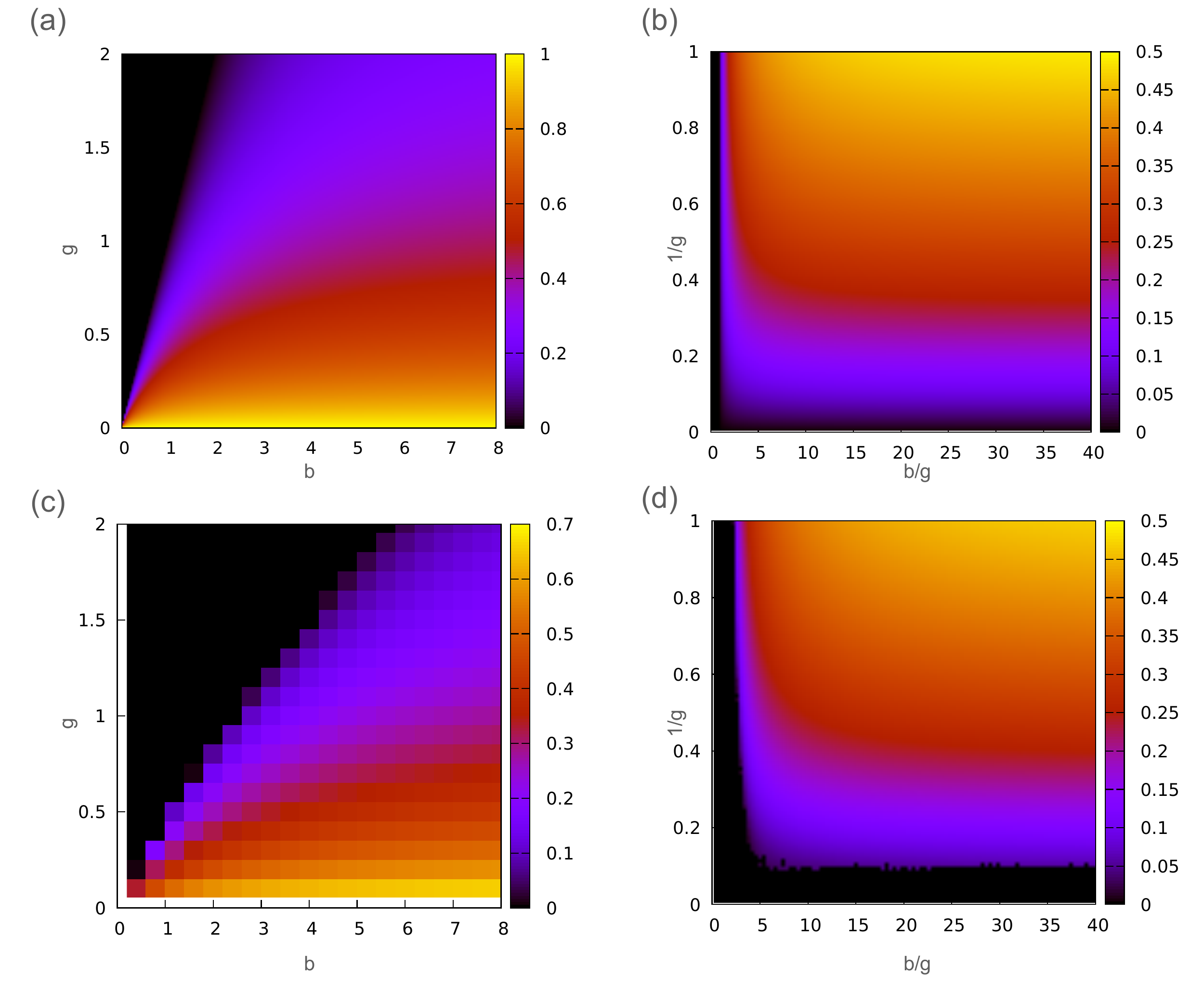}
%% Use \caption command for figure caption and label.
\caption{Plots of time-averaged  fraction of infected sites $I$  as a function of $b$ and $g$ in two different projections (see axes).
(a,b) from the mean field model.
(c,d) from the lattice model with $L=200$ and $L=100$ in (c) and in (d) respectively. Simulations  were performed by varying $\beta$ and $\gamma$ at a fixed value of $\delta=0.5$ (c), and by varying $\beta$ and $\delta$ at a fixed value of $\gamma=0.1$ (d). 
In all cases (a-d) we can see the transition between endemic ($\langle I\rangle\ne 0$) and non-endemic ($\langle I\rangle=0$) behaviour. In the mean field the transition occurs at $\beta=\gamma$. In the lattice model the transition line is close to $\beta= 2\gamma$, but bends for larger value of $\beta$, $\gamma$. %(b) Plot of the variation of $\langle(I-\langle I\rangle)^2\rangle$, which is system-size independent, showing that fluctuations and/or oscillations peak at the transition.   Simulations were run on a 100x100 grid: with low values of $I$ close to the boundary there is a risk of finite size effects removing all I sites.
%oscillation in the region close to the endemic threshhold.
}
\label{fig:I-time}
\end{figure*}

Having established the general nature of the behaviour, we investigate its dependence on the parameters.
In Figure~\ref{fig:I-time}, we show the long-time averaged fraction of infected sites in the system, $\langle I\rangle$, as a function of parameter ratios $\beta/\delta$ and $\gamma/\delta$ with $\delta$ setting the timescale, without loss of generality.
This shows that the dividing line between endemic and dying out, $\langle I\rangle=0$, is at $\gamma/\beta \sim 1$ for the mean-field and   $\gamma/\beta \sim 2$ for the lattice: the former being the epidemic threshold in SIR and the latter being the threshold for a wave to advance in two dimensional space~\cite{ackland2022fitting}. Despite the factor of two difference in onset, away from this threshold, the results are remarkably similar.  The lattice model in Figure~\ref{fig:I-time}d) also shows a region of zero long-term infections below $\delta=0.01$.  

\begin{figure*}[h]%
\centering
\includegraphics[width=\textwidth]{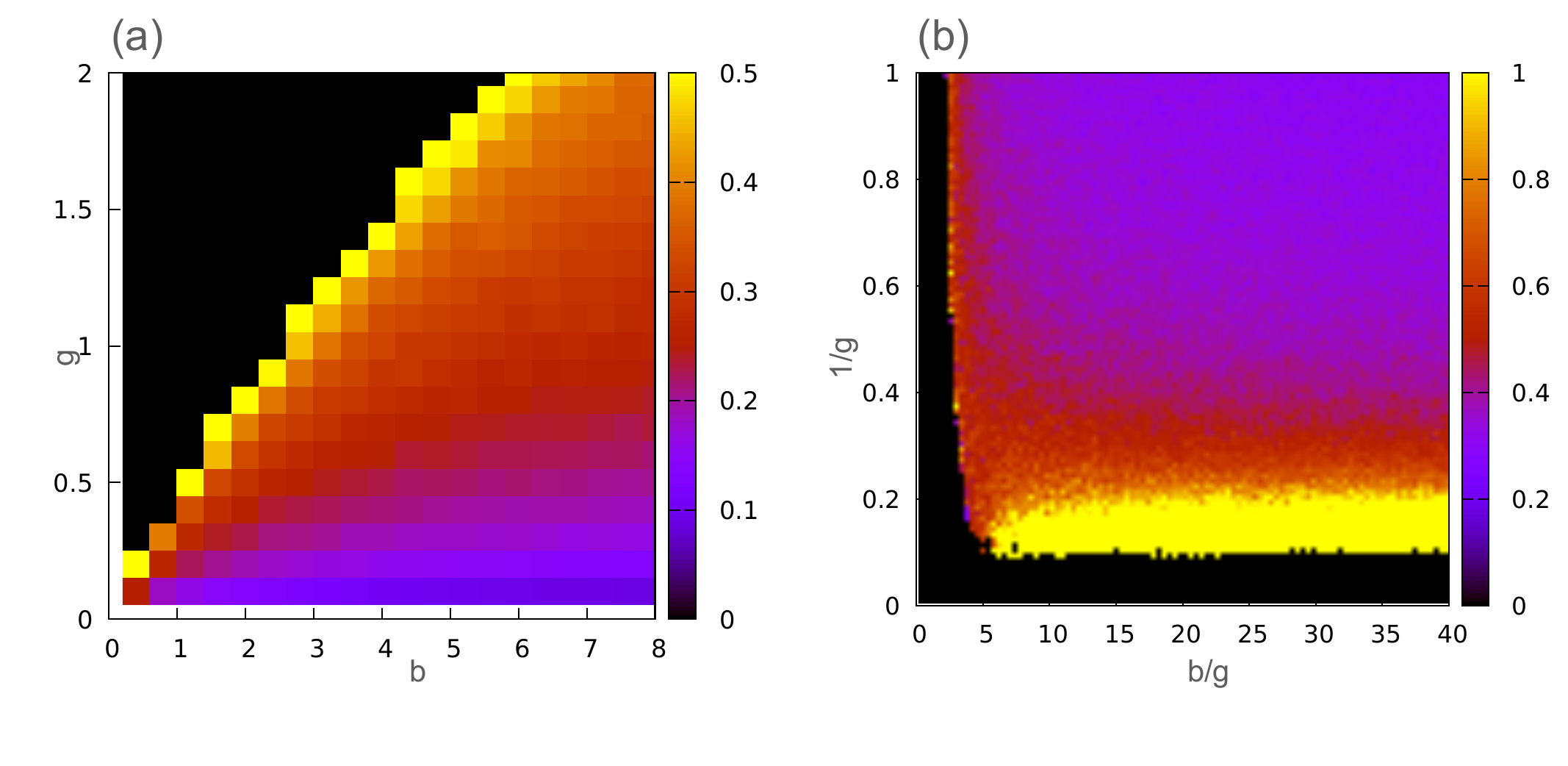}
\caption{Plots of the normalised variance of infected sites, 
$N\langle(I-\langle I\rangle)^2\rangle$, in the same planes as Fig.~\ref{fig:I-time} for the lattice model with (a) $L=200$, $\delta=0.5$ and (b) $L=100$, $\gamma=0.1$.  Inspection of the simulation trajectory reveals that regular oscillations are primarily observed in the high variance regime in the bottom right in panel (b), close to  $\delta/\gamma=1/g=0$, and hence close to the SIR limit. By contrast, the high normalised variance close to the epidemic threshold in (a) has no discernible period.}\label{fig:I-variance}
\end{figure*}

\begin{figure*}[t]%
\centering
\includegraphics[width=\textwidth]{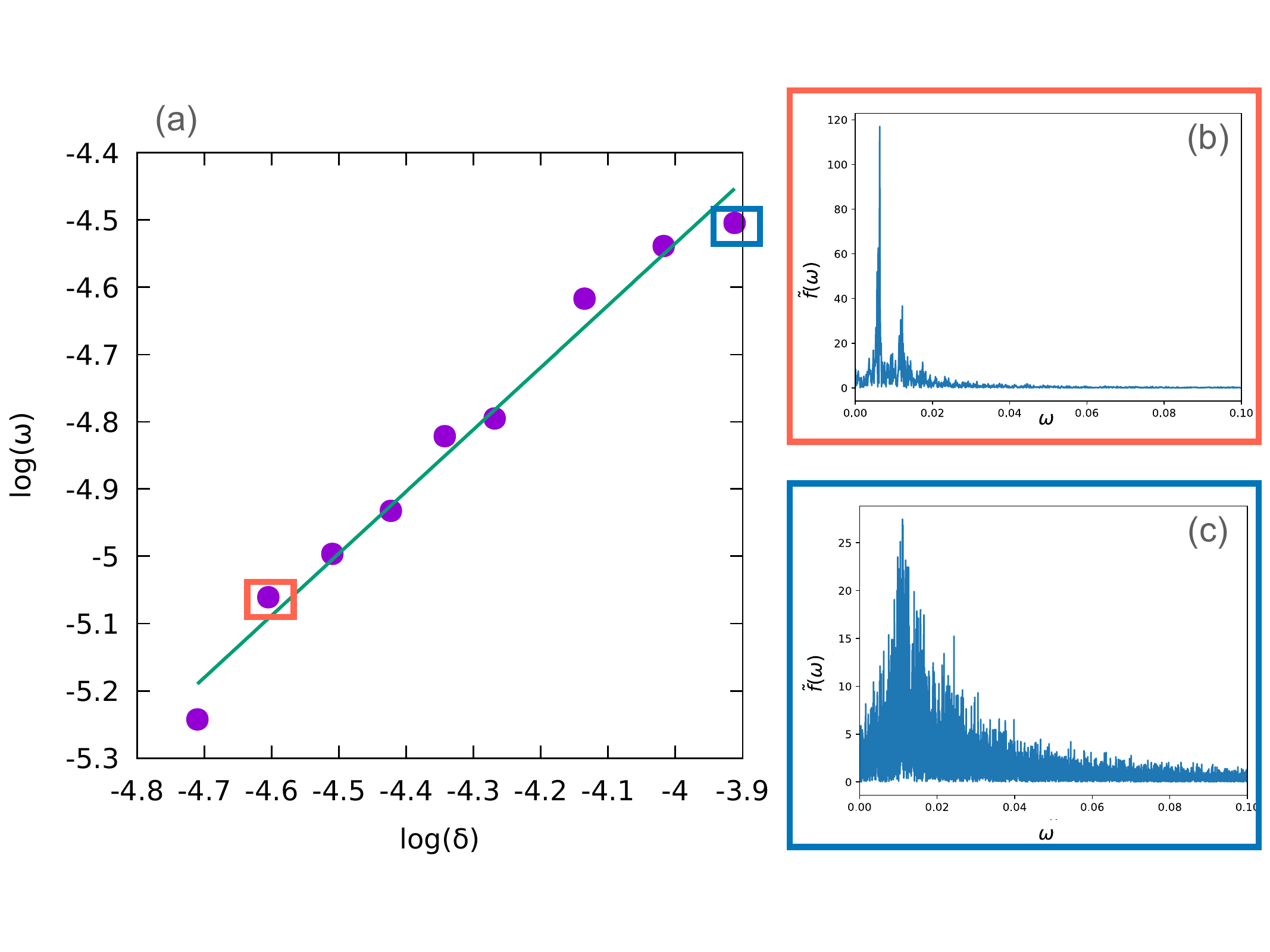}
\caption{Scaling of characteristic frequency in the SIRS lattice model with waning of immunity. (a) Typical frequencies  extracted from the peaks of Fourier transforms for selected time series showing oscillations for small $\delta$. The simulations were performed in a SIRS lattice model with $L=100$, the parameter range in the plot is located in the  bottom right region in Fig.~\ref{fig:I-variance}b showing high normalised variance of infected sites. The straight line is a fit corresponding to an exponent of $a\simeq 0.92$ (such that $\omega_{\rm max}\sim \delta^a$), compatible with linear behaviour. (b-c) Examples of numerical Fourier Transforms (b: $\beta=3.2$, $\gamma=0.1$, $\delta=0.01$; c: $\beta=3.2$, $\gamma=0.1$, $\delta=0.02$). Throughout this figure, frequencies are given in units of inverse simulation sweeps.} 
\label{fig:I-freq_lattice}
\end{figure*}

In Figure~\ref{fig:I-variance} we show the variance of this quantity $[\langle I^2 \rangle -\langle I\rangle^2]$ noting that, if measured as absolute number of sites, the variance scales with $N$ in the $N$-site SIRS lattice: the mean field model has no fluctuations and therefore zero variance.  Figure~\ref{fig:I-variance}a) shows large variations in infected numbers close to the epidemic threshold.  Fourier transforming the spectra reveals no characteristic frequency, so we can associate this phenomenon with the typical critical behaviour of physical systems approaching a phase transformation.  Figure~\ref{fig:I-variance}b) shows a different region of high variance as $\delta\rightarrow 0$, the limit in which SIRS becomes SIR.  Fourier transforming reveals that this variance does have a characteristic frequency, proportional to $\delta$ with a constant offset.  We can attribute this to ``boom and bust'' dynamics: the ``boom'' corresponds to the epidemic stage, during which the infection spreads across the system in a time inversely proportional to the wavespeed, which itself depends primarily on $\beta$\cite{fisher1937wave}.  The ``bust'' phase is the slow return to susceptibility until the system has lost herd immunity and the boom begins again.

It is notable that in both cases the oscillations occur most strongly in regions of lowest $\langle I\rangle$, meaning that in a finite system fluctuations are likely to reach $I=0$ at which point the infection is eradicated.

Interestingly, the lattice SIRS system evolves over time to have a characteristic lengthscale (see Fig.~\ref{graphicalabstract}).  This lengthscale tends to the lattice spacing as the separation of timescales between $\gamma$ and $\delta$ vanishes, at which point the characteristic frequency is lost.  Instead, at the thermodynamic limit, when the system size is much larger than the characteristic length, the characteristic frequency persists. 

\newpage
{\color{black}
\subsection{Spatial structures}

We observe a number of distinct spatial structures on the lattice model, depending on the parameters:

\subsubsection{Epidemic}
The simplest case is the early stages of an epidemic, where the infection spreads rapidly outwards from the initial site, forming a circular wave with R sites inside, and I sites at the interface.  This wave sweeps through the system, and dies out.

\subsubsection{Endemic}
The late stages of the non-oscillating endemic are characterised by coexisting SIR sites with no particular emergent structures.

\subsubsection{Oscillating}
The oscillating endemic state shows emergent structures at a characteristic scale, which can be regarded as flare-ups of infection.  Since the system shows oscillation in the total number of infecteds, these outbreaks are correlated.  

\subsubsection{Contained outbreak}
Below the SIR epidemic threshold, the initial I sites spawn finite-sized outbreaks which die out. This state is measurably distinct from the Epidemic case in the scaling with system size: The total number of infections per initial I is a constant, independent of system size, whereas the epidemic infects a finite fraction of sites.
the size of the outbreaks is dependent on details of the model. 
}

\begin{figure*}
    \includegraphics[width=0.49\textwidth]{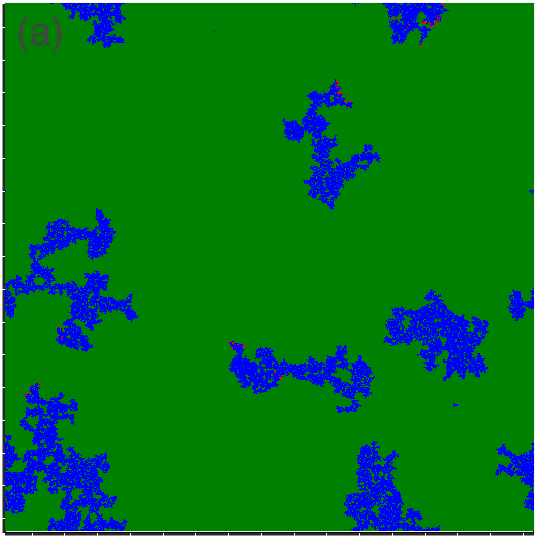}
    \includegraphics[width=0.49\textwidth]{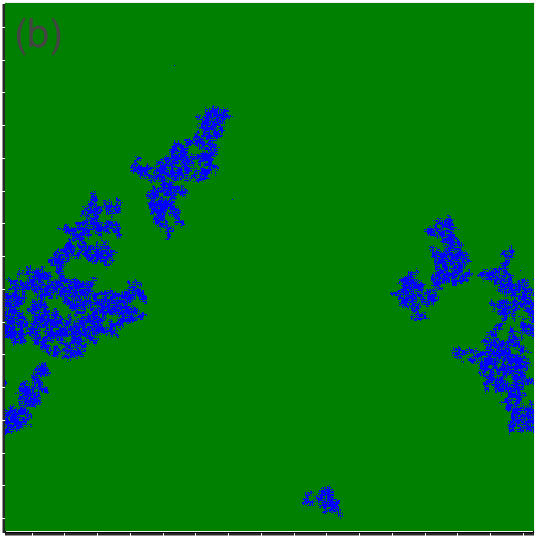}
    \caption{Contained outbreaks close to the epidemic threshold from 10 randomly placed initial cases.  (a) Near-neighbour only interactions ($\beta_a=0.4,\gamma=0.1$
    (b) First and second-neighbours $\beta_a=0.275,\gamma=0.1$,  We set $\delta=0$ so the full extent of the outbreak is revealed.
\label{fig:animals}
}\end{figure*}

{\color{black}
\subsection{Algorithm Dependence}

There are different possibilities for the final step in the update algorithm, for example, we might consider an 8-site neighbourhood which allows infection two sites away, with mean-field equivalent probability $\beta_a/8$.  This is particularly important at the epidemic threshold.   
As an example, Figure~\ref{fig:animals} shows contained outbreaks just below the epidemic threshold from the two algorithms.  The first observation is that the 8-neighbour has a 45$\%$ lower epidemic threshold value of $\beta_a$.  The  outbreak structures are completely different from the epidemic case (Fig.~\ref{fig:I-time}) and rather similar to each other.  The 8-neighbour can be seen to be somewhat more filamentary and less compactified, but the spatial shapes of the outbreak structures are not so different that an easy explanation of the shift presents itself.  Some of the effect may come from the fact that, in the 4-neighbour case a maximum of 3/4 neighbours of a newly infected site are susceptible, whereas for the 8-neighbor case the maximum is 7/8, and the longer range can be considered a first step towards the mean-field $\beta_a=\gamma$.
}

\subsection{Scaling of oscillation timescale}

If we assume that a wave of infection spreads through the system and replaces all susceptibles with infecteds immediately, then at the top of the peak the infected population will be equal to \(S^*(\beta,~\gamma)\), the critical susceptible fraction for herd immunity. %, \(S^*=2g/b\) on our square lattice. 
The subsequent dynamics will be \(I=S^*e^{-\gamma t}\). Assuming that almost all the infected will have died out before a significant fraction of \(S\) return, the dynamics of \(S\) and \(R\) will be
\begin{eqnarray}
\frac{dS}{dt} & = & \delta R \,,\\
\frac{dR}{dt} & = & \gamma I - \delta R \,,
\end{eqnarray}
whose solution is
\begin{eqnarray}
S & = & 1-\frac{\left(S^*+g-1\right)e^{-\delta t}-S^*e^{-\gamma t}}{g-1} \,,\\
R & = & \left(\frac{g-1+S^*}{g-1}\right)e^{-\delta t}-\left(\frac{gS^*}{g-1}\right)e^{-\gamma t}\,. \label{SR_sols_after_peaks}
\end{eqnarray}
We predict that the time between peaks will be approximately given by the time $t^*$ required for \(S\) to rise above \(S^*\), i.e., from Eq.~\ref{SR_sols_after_peaks}, \(S(t^*)=S^*\). In the $g\gg 1$ limit, i.e., where the timescale $1/\delta$ is limiting, the solution is \[t^*=\delta^{-1}\ln\left(\frac{1}{1-S^*}\right)\,.\] Hence, we predict that the frequency of oscillations will scale with \(\delta\). For $S^*\sim 1$, i.e., close to the endemic threshold, there will be a logarithmic divergence in the time scale. Far away from the transition, $S^*\ll 1$, and we will have $t^*=S^*/\delta$.  %=2\gamma/\beta\delta$. 

This scaling behaviour is consistent with the simulations in Figure~\ref{fig:I-freq_lattice} and in sharp contrast with the eigenvector analysis in Figure~\ref{fig:I-freq}. Relaxing the initial assumption that the infection is instantaneous ($\beta\rightarrow\infty$) will introduce a $\beta$-dependence, increasing the period by approximately the time taken for the wave to spread.

\section{Discussion and Conclusions}

We have compared three variants of the two-parameter SIRS model with a view to determining whether it can reproduce the kind of oscillations seen in the endemic state of omicron SARS-CoV-2.

The standard ODE-based mean field model supports only a steady endemic state at long time, reached by either exponential decay, or by damped oscillations.  This behavior was studied using an eigenvalue analysis revealing three regions:  non-endemic, overdamped endemic with negative real eigenvalues and underdamped endemic with complex eigenvalues.  

The stochastic ODE model exhibits persistent oscillations whenever there is an imaginary part to the eigenvalue.  Larger eigenvalues correspond to larger stochastic oscillation frequencies, however the relationship is not straightforward. The stochastic frequency depends on the size of the noise applied, indicating that the oscillation is anharmonic.  For epidemics, the slowest process is typically loss of immunity, $\delta\rightarrow 0$, and in this case the stochastic ODE has oscillations with frequency $\sim \sqrt{\delta}$.

\begin{figure*}
    \includegraphics[width=\textwidth]{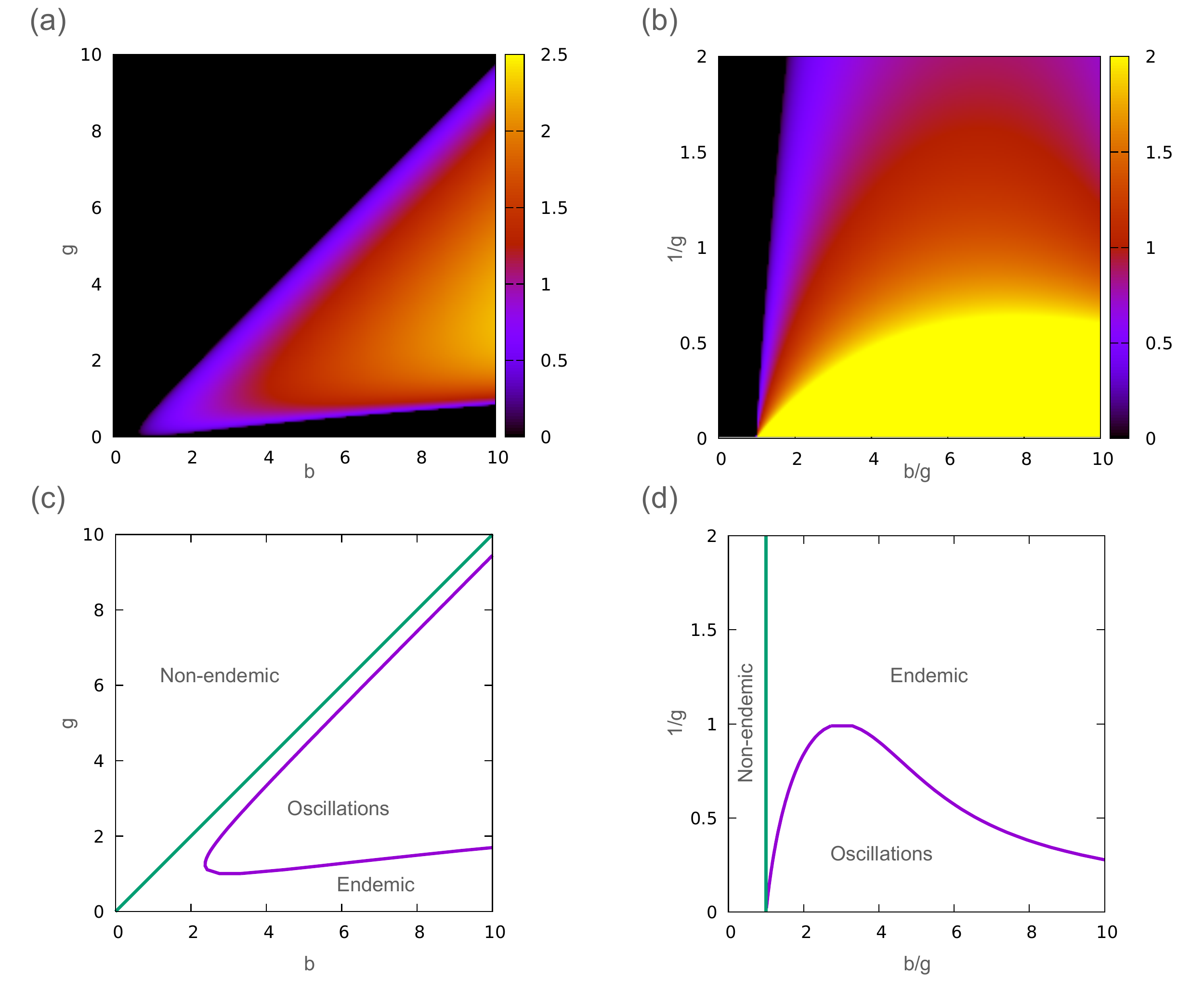}
    \caption{
    Mean field phase diagrams showing non-endemic, endemic and oscillations regimes. (a) Imaginary part of the complex eigenvalue $\lambda$, or frequency, from Eq.~\ref{eq:evals}, in the $(b,g)$ plane. The frequency is in units of $\delta$ (as we set $\delta=1$ in Eq.~\ref{eq:evals}). (b) Imaginary part of the complex eigenvalue, or frequency, in the $(b/g,1/g)$ plane, again in units of $\delta$ (the colour range has been cut to ease visualisation of the pattern. %underscoring dependence on $\delta$. The frequency is in units of $\delta$. %-- the normalisation bas been changed with respect to (a) to aid visualisation of the pattern where the imaginary part is non-zero.
    (c,d) Mean field phase diagrams in the $(b,g)$ plane (c), and in the $(b/g,1/g)$ plane (d). The oscillation regime is defined to be the one where the ratio between the absolute values of the imaginary and the real parts of the eigenvalue $\lambda$ is $\ge 1$, which corresponds to sustained and visible underdamped oscillations. %where $\lambda$ is the eigenvalue describing the decay of fluctuations around the endemic state (see text). 
    Note that $1/g=0$ corresponds to $\delta=0$ and the SIR model.\label{fig:phasediagram_MF} 
    }
\end{figure*}

The lattice SIRS model is dramatically different from the ODE-based model.  Firstly, the epidemic threshold is at $\beta/\gamma=2$, not at the conventional SIR $\mathcal{R}_0=\beta/\gamma=1$.
This has previously been observed in the SIR model and can be attributed to the nature of wavelike growth.  Secondly, there are large fluctuations  in the regions of parameter space closest to the epidemic threshold - precisely the opposite behaviour to the mean field models which have a non-oscillating region (Fig.~\ref{fig:phasediagram_MF}) at the  epidemic threshold.  

The different epidemic thresholds  can be attributed to the correlation between the location of S and I sites.  In the lattice model, each newly infected site must be adjacent to at least one previously infected site, i.e. at least 25\% of its neighbours are I, and those neighbours must have 50\% of their sites either I or recently I.  Comparing this with the typical mean-field case which, initially, implies that essentially all ``neighbours''  are S.  This reduction of IS-neighbour pairs due to correlation is responsible for the shift in the epidemic threshold from $\beta/\gamma=1$ to $\beta/\gamma=2$ whenever the system is big enough for the outbreak to form a wave of advance.   This explains why the $\delta$ dependence is seen only at low $\delta$ - if the R site becomes S ``quickly'' (i.e. on the timescale of the wave speed) then this strong correlation is broken and, indeed, the periodic oscillation also disappears.

In the lattice model, it appears that a characteristic lengthscale emerges over time as the time-averaged $\langle I \rangle$ increases.  We speculate that the system is self-organising to the ``maximum life'' steady state in which the number of replicators - in this case the I site - is maximised\cite{ackland2004maximization,ackland2004stabilization,wood2008daisyworld,mitchell2009boom,mitchell2007strategy,edmonds2009towards,hellier2018diffusion}, similar to the principle of maximum entropy production\cite{martyushev2006maximum,ozawa2003second,dewar2005maximum,kleidon2010maximum}.

{\color{black} The details of the lattice model waves are somewhat sensitive to the algorithm, in which infection is independent of how many adjacent sites are I.  The infection can spread only if each I site infects at least one neighbour, on average. Each infected site is surrounded by between 0 and 3 susceptibles, leading to fastest growth occuring at more isolated regions of an outbreak. Close to the epidemic threshold, only spindly structures with growth at their tips can grow, which gives a shape to failed outbreaks.  This is quite distinct from the circles seen for the epidemic case, where a flat interface can extend as a wave.  By contrast, the oscillations are a robust feature in both mean-field and lattice, and therefore we can expect them to be independent of the lattice type. Rewiring\cite{ackland2022fitting} the network connections takes the model from a lattice through a small-world model to an Erdos-Renyi Graph.  A key issue here is whether enough near-neighbour connectivity survives to support a wave.  However the oscillations depend mainly on the rate of immunity loss by individuals ($\delta$), and so will be present regardless of network size.}

In summary, we have shown that the SIRS model with two independent parameters can give rise to spontaneous oscillations in the number of infected individuals due to either stochastic noise in a well-mixed population or spatial effects %separation 
in a lattice model. 
The  oscillations can be characterized by boom and bust dynamics, where sharp epidemic phases (timescales set by $\beta$ and $\gamma$)  are separated by low-$I$ periods of waning immunity (timescale set by $\delta)$.  

The work emphasizes that epidemics can behave completely differently depending on details of the spreading process: either exponential or wavelike.  In particular, that successful suppression, delay and predictability  can be achieved by forcing interactions to remain localised to enforce a wavelike outbreak.  The fact that throughout the COVID epidemic $\mathcal{R}\sim 1$, and the subsequent oscillation in $I(t)$ once restrictions were removed suggests that SIRS is the appropriate
model.  It can be inferred that the loss of immunity  to the SARS-Cov2 omicron variant, whether due to viral evolution or changes in the host immune system, is approximately ten weeks.  

{\color{black}
\subsection{Practical consequences}
%We originally declined to speculate on the implications of this study for Covid-19 or future pandemics.  This section is added at the behest of the referees.

We end this study with a discussion of the possible implications of this study for Covid-19 and future pandemics, emphasizing the necessarily speculative nature of this discussion.

The study is an idealised model, but has a number of implications for disease control in future pandemics.  Firstly, the fact that the number of cases drops after a certain time may be due to the dynamics of the disease itself, unrelated to any mitigating effects applied. 
Also, such a drop should not necessarily be interpreted as the ``end'' of the epidemic. Many assumptions are required to disentangle these effect from the pandemic data, such that reasonable-looking assumptions can assign the end of each wave of Covid either to interventions, variants or natural oscillations.  For scientific study, data from an intervention-free period is essential, and it is unfortunate that the cessation of detailed data collection coincided with the end of interventions.
}

{\color{black} The network structure is extremely important, and our study represents two opposite limits.  The ODE models implicitly assume well-mixed population, while the lattice model represents a two-dimensional space with no long-ranged connections.  The spread is completely different in each case.  For a start, the critical value of $\mathcal{R}_0$ leading to an epidemic goes from $\mathcal{R}_0=1$ in the well mixed case to 
$\mathcal{R}_0=2$ for the lattice \cite{ackland2022fitting}.  Also, the lattice epidemic is wavelike, not exponential, so $R(t)$ will remain close to 1 \cite{ackland2022fitting} and is not a useful measure for driving policy decisions. Interestingly, interventions such as travel bans work by making the network more spatially localised, moving along the continuum between the well mixed and local-lattice extremes.

There is a significant difference between oscillations in endemic Covid and influenza: in the latter case it is long established that there is a ``flu season'' in the winter, in both Northern and Southern hemispheres(and twice a year in Singapore).  There is no obvious periodic driver for COVID infections, and this work demonstrates that such a driver is not necessary.

A further important result is that the spreading algorithm matters. As demonstrated in both mean-field and non-nearest neighbour cases, for the same number of infections per initial case ($\mathcal{R_0}$) the epidemic threshold is much lower when those infections are to more remote contacts. 
%Clearly the ``remote'' contacts are more important in driving the epidemic than ``near-neighbour'' ones, even when accounting for number of contacts{I don't get this: if you made it that infection was \textit{only} possible to next-nearest neighbours, I think this would result in a similar threshold to the nearest-neighbour-only case. AB}. 
%The longer-ranged spread can be associated with airbourne respiratory viruses, which may explain the ineffectiveness of short-ranged preventative measures such as masks and handwashing in the recent COVID epidemic~{I think this is getting a bit too speculative. I would delete this last sentence. AB}.
 \cite{jefferson2023physical}
}

GJA would like to thank UKRI and EPSRC for support under grants EP/V053507/1 and ST/V00221X/1.

%\clearpage 

%\begin{figure}[h]%
%\centering%% For centre alignment of image.
%\includegraphics{Figure2.pdf}
%\caption{Plot of the fraction of sites ever infected as a function of  $\beta/\delta$ and $\gamma/\delta$ for $N=200$, as in Fig.~\ref{fig:I-time}.  All sites are infected in the endemic case, the epidemic case (close to the transition) %$\mathcal{R}_0\rangle2$ infects a fixed fraction of the sites, while the non-epidemic case $\mathcal{R}_0\langle2$ infects a fixed number of sites}\label{fig:I-number}
%\end{figure}

%\bibliography{WSS}


%\end{thebibliography}
\end{document}